\documentclass[12pt]{article}
\usepackage{english}
\usepackage{a4}
\usepackage{epsf}
\usepackage{epsfig}

\begin{document}
\bibliographystyle{unsrt}
\noindent
\begin{center}
{\Large \bf Growing length scales in a supercooled liquid close to an
interface}

\vspace*{5mm}

{\large 
Peter Scheidler$^*$, Walter Kob$^\ddagger$\footnote{Author to whom 
correspondence should be addressed}, Kurt Binder$^*$, and Giorgio
Parisi$^\#$}\\

$^*$ Institut f\"ur  Physik, Johannes Gutenberg-Universit\"at, 
Staudinger Weg 7, D-55099 Mainz, Germany\\

$\ddagger$ Laboratoire des Verres, Universit\'e Montpellier II, F-34095
Montpellier, France\\

$^\#$Dip. di Fisica, Universit\`a di Roma ``La Sapienza'', INFM and INFN,
P.le A. Moro 2, I-00185 Roma, Italy

\end{center}

\vspace*{7mm}
\par
\noindent
\begin{center}
\begin{minipage}[h]{122mm}
We present the results of molecular dynamics computer simulations
of a simple glass former close to an interface between the liquid
and the frozen amorphous phase of the same material. By
investigating $F_s({\bf q},z,t)$, the incoherent intermediate scattering
function for particles that have a distance $z$ from the wall, we
show that the relaxation dynamics of the particles close to the wall
is much slower than the one for particles far away from the wall. For
small $z$ the typical relaxation time for $F_s({\bf q},z,t)$ increases
like $\exp(\Delta/(z-z_p))$, where $\Delta$ and $z_p$ are constants. We
use the location of the crossover from this law to the bulk behavior
to define a first length scale $\tilde{z}$.  A different length scale
is defined by considering the Ansatz $F_s({\bf q},z,t)=F_s^{\rm bulk}({\bf
q},t)+a(t)\exp\left[-(z/\xi(t))^{\beta(t)}\right]$, where $a(t)$, $\xi(t)$, and
$\beta(t)$ are fit parameters. We show that this Ansatz gives a very
good description of the data for all times and all values of $z$. The
length $\xi(t)$ increases for short and intermediate times and decreases
again on the time scale of the $\alpha$-relaxation of the system. The
maximum value of $\xi(t)$ can thus be defined as a new length scale
$\xi_{\rm max}$. We find that $\tilde{z}$ as well as $\xi_{\rm max}$
increase with decreasing temperature. The temperature dependence of
this increase is compatible with a divergence of the length scale at
the Kauzmann temperature of the bulk system.

\end{minipage}
\end{center}

\vspace*{5mm}
\par
\noindent

\section{Introduction}
The idea that the slow dynamics of supercooled liquids is related in some
way to the existence of domains in which the dynamics of the particles is
cooperative is an old one (Adam and Gibbs 1958). The size
of such domains is supposed to grow with decreasing temperature and can
in turn be used to explain the slowing down of the dynamics. For a long
time it was not clear at all whether or not the idea of such domains has
any counterpart in real glass formers or whether it is just an appealing
theoretical concept. Only in recent years a variety of novel experimental
techniques yielded results that gave evidence that such domains do indeed
exist in the supercooled state (e.g.  Cicerone {\it et al.} 1995, Ediger
2000, Hempel {\it et al.} 2000, Leheny {\it et al.} 1996, Richert 1996,
Sillescu 1999, Tracht {\it et al.} 1998, Yamamuro {\it et al.} 1998). 
Also computer simulations of two
and three dimensional model systems give evidence that in the supercooled
state the relaxation dynamics of the particles is very cooperative (Doliwa
and Heuer 2000, Donati {\it et al.} 1999a/b, Perera and Harrowell 1999,
Yamamoto and Onuki 1998).  Progress has also been made on the theoretical
level in that it was shown that such domains are related to the nonlinear
susceptibility of a certain four point correlation function (Franz and
Parisi 1998, Franz {\it et al.} 1999, Franz and Parisi 2000).

In the present paper we use the fact that the relaxation dynamics of
a particle in a confined system depends strongly on its distance $z$
from the wall (Scheidler {\it et al.} 2000, Scheidler 2001). 
By investigating how this
influence depends on $z$ and time $t$ we thus can determine a dynamical
length scale and study its dependence on temperature.

\section{Model and Details of the Simulations}

The system we study is a binary (80:20) mixture of Lennard-Jones
particles, i.e. with an interaction potential $V_{\alpha\beta}(r)
= 4\epsilon_{\alpha\beta} [(\sigma_{\alpha\beta}/r)^{12} -
(\sigma_{\alpha\beta}/r)^6]$. This potential was truncated and shifted at
2.5$\sigma_{\alpha\beta}$. The interaction parameters are $\sigma_{\rm
AA}=1.0$, $\epsilon_{\rm AA}=1.0$, $\sigma_{\rm AB}=0.8$, $\epsilon_{\rm
AB}=1.5$, $\sigma_{\rm BB}=0.88$ and $\epsilon_{\rm BB}=0.5$, where A and
B label the type of particle. In the following we will discuss the results
in terms of reduced units, using $\sigma_{\rm AA}$ and $\epsilon_{\rm
AA}$ as the units of length and energy (setting the Boltzmann constant
$k_B=1$), and  $(m \sigma_{AA}^2/48 \epsilon_{AA})^{1/2}$ as the unit
of time. (Here $m$ is the mass of the particles.)

In the past the static and dynamical properties of this system {\it in
the bulk} have been analyzed carefully (Kob and Andersen 1995a/b, Gleim
{\it et al.} 1998) and it was found that its dynamics is described very
well by means of the so-called mode-coupling theory (G\"otze 1999) with a
critical temperature around $T_c=0.435$. Thus this is the temperature at
which the dynamics of the {\it bulk} system slows down significantly. A
further relevant temperature for this system is the Kauzmann temperature
$T_K$ which has been estimated to be around 0.3 (Coluzzi {\it et al.}
2000a/b, Sciortino {\it et al.} 1999).

The confined liquid studied in the present work was set up in the
following way: We first equilibrated a system of 3000 particles in a box
of dimensions $L_x=L_y=12.8$ and $L_z=15$ by applying periodic boundary
conditions in all three directions. After equilibration we joined a copy
of the configuration in the positive $z$-direction and one in the negative
$z$-direction. The particles of these two copies that were in the layer
(perpendicular to the $z$-axis) of thickness 2.5 closest to the original
system were subsequently frozen and constituted the wall for the confined
system (2.5 is the range of the interaction). In this way we thus 
generated a sandwich geometry in which a
fluid system of extension $L_x\times L_y \times L_z$ was confined by
two walls of dimension $L_x\times L_y\times 2.5$.  (In the direction of
the $x$ and $y$ axis we kept the periodic boundary conditions.) 
The width $L_z$ of the film was sufficient to make sure that the dynamical bulk
properties are realized in the center. Apart
from the two mentioned walls we also added a hard-core potential at
$z=0$ and $z=L_z$ to avoid a penetration of particles into the wall. 
It is now straightforward to show that with these
types of boundary conditions the equilibrium structure of the confined
liquid is {\it identical} to the one of the bulk liquid. Hence it is not
necessary to equilibrate the confined system after the introduction of the
wall. (Note that this procedure was done for all temperatures investigated
and hence the structure of the wall depends slightly on temperature.) 
Moreover the energy of a particle near to the wall is statistically
the same to the one of a particle far from the wall: in general the presence of
such a wall cannot be discovered by looking to quantity measured at one
time, but influences only the correlation of quantities measured at two
different times.

The
lowest temperature investigated was $T=0.5$ for which production runs of
16 million time steps were used. Note that the length of these production
runs is more than ten times longer than the one needed in a simulation
for the bulk (Kob and Andersen 1995a/b), since the relaxation of the system close to the
wall is so much slower. Finally we mention that in order to improve the
statistics of the results we averaged them over 8-16 independent runs.

\section{Results}
\begin{figure}[htb]
\psfig{file=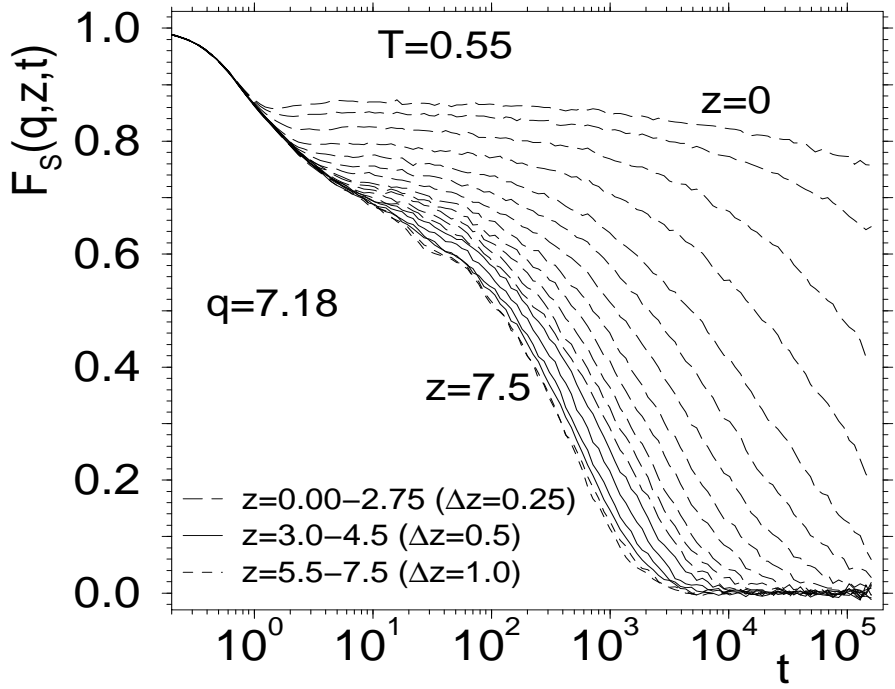,width=12cm,height=9cm}
\vspace*{-4mm}

\caption{Time dependence of the generalized incoherent intermediate
scattering function from equation~(1) for the A particles for $T=0.55$. 
The curves are
for $q=7.18$, the location of the main peak in the static structure
factor. The different curves correspond to different distances $z$
from the wall (see label of curves).}
\label{fig1}
\end{figure}

The dynamical quantity that we investigate is the incoherent intermediate
scattering function $F_s({\bf q},t)$, i.e. a space-time correlation
function that is important from the theoretical as well as experimental
point of view (Hansen and McDonald 1986). Due to the presence of the
walls the system is no longer homogeneous or isotropic and hence we
have to consider the $z$-dependence of $F_s({\bf q},t)$. For this we
investigate the following generalization of the incoherent intermediate
scattering function:
\begin{equation}
F_s({\bf q},z,t) = \frac{1}{N_{\alpha}} \left\langle
\sum_{j=1}^{N_{\alpha}}
\exp(i{\bf q} \cdot ({\bf r}_j(t)-{\bf r}_j(0)))
\delta(z_j(0)-z)\right\rangle
\label{eq1}
\end{equation}
where the $\delta$-function on the right hand side makes sure that 
one considers only particles that at time $t=0$ were a distance $z$
away from the wall. The wave-vectors $\bf q$ we consider in the following
are parallel to the walls and their $x$ and $y$ components are chosen
such that they are compatible with the periodic boundary conditions.
In Fig.~\ref{fig1} we show the time dependence of $F_s({\bf q},z,t)$
for various values of $z$. From this figure we see that the relaxation
dynamics of the particles close to the walls (curves for $z\approx 0$)
is much slower than that for particles in the middle of the film (curves
for $z\approx 7.5$). In an earlier paper we have shown that the typical
relaxation time for particles that have a distance $z$ from the wall,
$\tau(z,T)$, increases like 
\begin{equation}
\tau(z,T) \propto \exp \left[ \Delta/(z-z_p) \right], 
\label{eq2}
\end{equation}
where $\Delta\approx 7.5$ and $z_p\approx -0.5$ are constants and the
proportionality factor depends on $T$ (Scheidler
{\it et al.} 2000, Scheidler 2001). This law is valid for small values
of $z$, i.e.  close to the wall. For large $z$ we recover instead the
bulk relaxation time $\tau_{\rm bulk}(T)$. The distance from the wall at
which one finds the crossover from the law given by equation~(\ref{eq2})
to the constant bulk value $\tau_{\rm bulk}(T)$ can be used to define
a length scale $\tilde{z}$ (Scheidler 2001). 
Thus $\tilde{z}$ is the length scale over
which the wall influences the dynamics of the particles and below we
will discuss how this scale depends on temperature.

\begin{figure}[htb]
\psfig{file=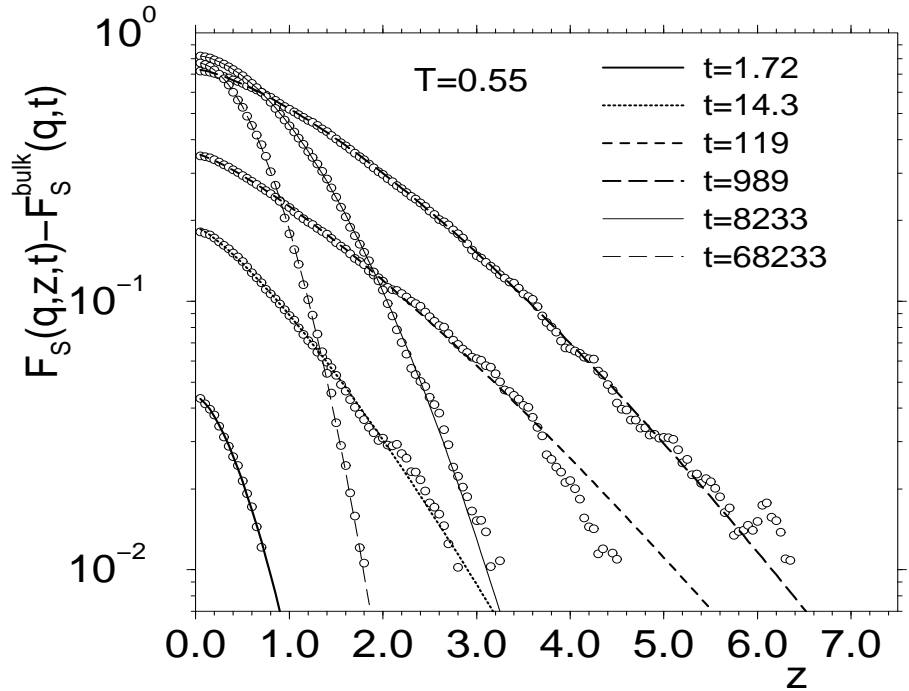,width=12cm,height=9cm}
\vspace*{-4mm}

\caption{The time dependence of $F_s({\bf q},z,t)-F_s^{\rm bulk}({\bf
q},t)$ is shown to check
the validity of the Ansatz given by equation~(3) (the value of $q$ is the
same as in figure 1). The symbols are the data from the
simulations and the smooth curves are the fits with the functional form given by
equation~(3).}
\label{fig2}
\end{figure}

A second approach to investigate the spatial dependence of the structural 
relaxation is to consider the whole $t$- and $z$-dependence of
the intermediate scattering function within one Ansatz. From 
figure~\ref{fig2} we see that the $z$-dependence of $F_s({\bf
q},z,t)$ is very smooth. In particular it is clear that for $z \gg 1$ this
function has to converge to the intermediate scattering function for the
bulk, $F_s^{\rm bulk}({\bf q},t)$. Therefore it is reasonable to make
an Ansatz of the form
\begin{equation}
F_s({\bf q},z,t)=F_s^{\rm bulk}({\bf q},t)+
a(t) \exp \left[-(z/\xi(t))^{\beta(t)} \right],
\label{eq3}
\end{equation}
i.e. that the whole $z$-dependence of $F_s({\bf q},z,t)$ is given by a
stretched exponential. Here $a(t)$, $\xi(t)$, and $\beta(t)$ are unknown
functions of time that can be found from fitting at a given time the
$z$-dependence of $F_s({\bf q},z,t)$ with the functional form given by
equation~(\ref{eq3}). Note that $F_s^{\rm bulk}({\bf q},t)$ is known from
the simulations of the bulk and is hence not a fit parameter.

That the Ansatz~(\ref{eq3}) is indeed able to describe the $z$-dependence
of $F_s({\bf q},z,t)$ very well is demonstrated in figure~\ref{fig2}
where we show $F_s({\bf q},z,t)-F_s^{\rm bulk}({\bf q},t)$ as a function
of $z$ for times that span the range from microscopic times to
the time of the $\alpha$-relaxation, i.e. of the final decay of the
correlation functions. We also mention that other simple functional forms,
e.g. if the stretching parameter $\beta$ is set to 1.0, do not give
satisfactory fits (Scheidler 2001). Nevertheless such a purely exponential
Ansatz leads to a very similar dynamical length scale.

\begin{figure}[htb]
\psfig{file=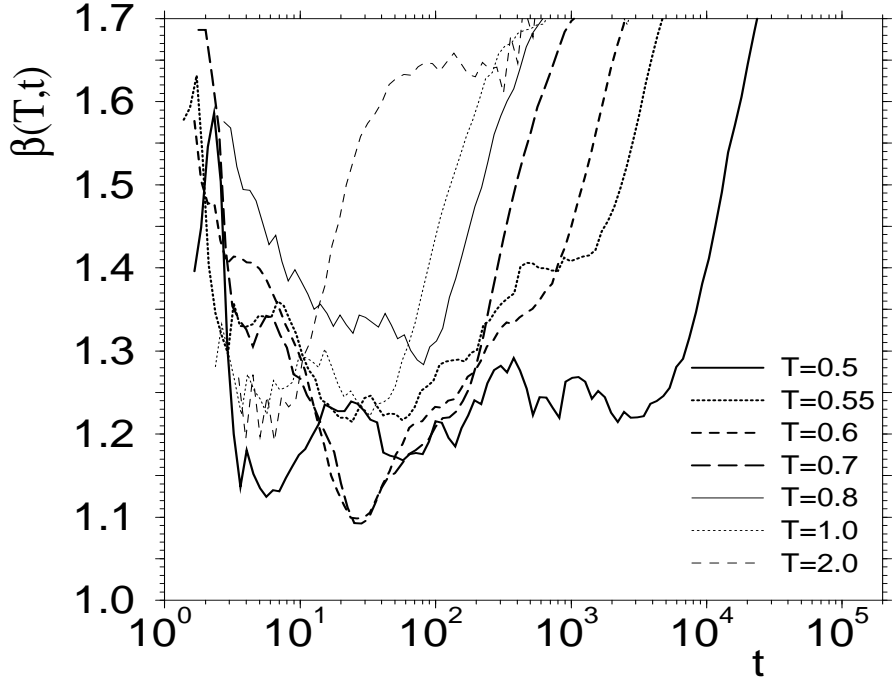,width=12cm,height=9cm}
\caption{Time dependence of the stretching parameter $\beta(t)$ from equation~(3) for
all temperatures investigated.}
\label{fig3}
\end{figure}

In figure~\ref{fig3} we show the time dependence of the stretching
parameter $\beta(t)$ for all temperatures investigated. From this figure
we recognize that at low temperatures $\beta(t)$ shows at short times
($\approx 3$ time units) a rapid decrease from values around 1.5 to
a value around 1.2. Subsequently it remains at this latter value for
a time span that corresponds roughly to the $\beta$-relaxation, i.e.
the time window during which the correlation functions for small $z$
remain close to the plateau and which extends at the lowest temperature
over roughly three decades in time (see figure~\ref{fig1}). For longer
times $\beta(t)$ increases again to values above 1.4. For intermediate and
high temperatures the time dependence of $\beta(t)$ remains qualitatively
the same as the one  we just described. The main difference is that the
length of the plateau at intermediate times is shortened significantly,
in agreement with the fact that also the correlation functions show only
a short plateau at high temperatures (Scheidler 2001). Note that the fact
that at intermediate times $\beta(t)$ is larger than 1.0 shows that the
$z$-dependence of the $F_s({\bf q},z,t)$ is rather strong, i.e. it is not
just an exponential dependence.  Additionally we observed in {\it thin}
films that the influence of two walls on the dynamics of the particles
between them is not just the superposition of two laws of the form given
by equation~(\ref{eq3}).  This is thus evidence that the slowing down of
the dynamics is related to the presence of a strongly non-linear process
such as, e.g., the non-linear feedback process of mode-coupling theory.

\begin{figure}[htb]
\psfig{file=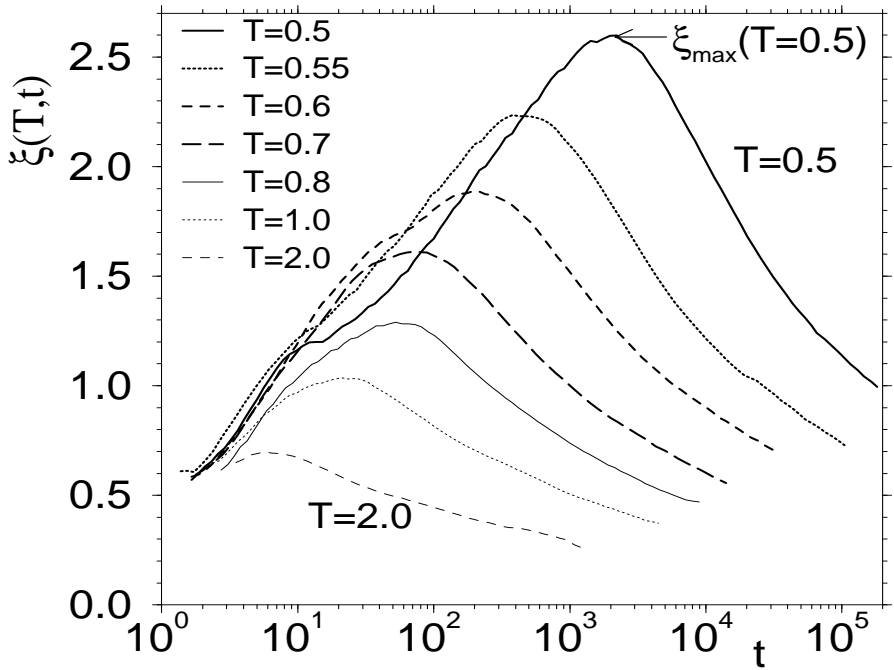,width=12cm,height=9cm}
\caption{Time dependence of the scale $\xi(t)$ from equation~(3) for
all temperatures investigated. The value of the maximum of the curves is used to
define the length scale $\xi_{\rm max}$.}
\label{fig4}
\end{figure}

In figure~\ref{fig4} we show the time dependence of the parameter $\xi(t)$
for all temperatures investigated. We see that for short and intermediate
times this length scale increases, attains a maximum around a time
that corresponds to the $\alpha$-relaxation time of the bulk system,
and then starts to decrease again. The value of $\xi$ at its maximum
can thus be used to define a new length scale $\xi_{\rm max}(T)$.
Note that the time at which this maximum is attained is roughly in the
time window in which $\beta(t)$ has a minimum, i.e. is relatively small.
This implies that at this time the influence of the wall on the dynamics
of the particles in the fluid extends over the largest range.

\begin{figure}[tb]
\psfig{file=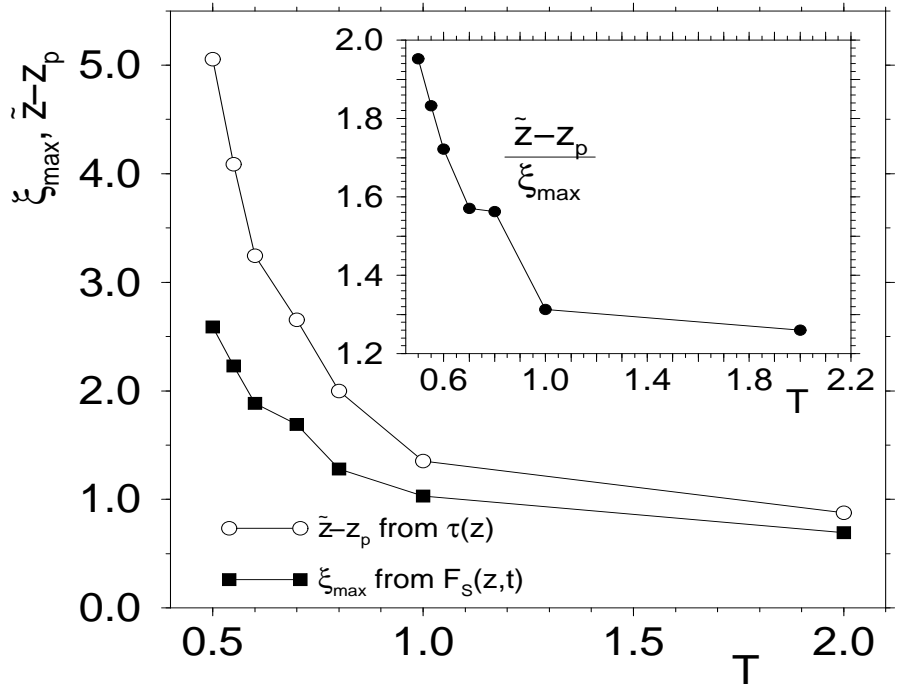,width=12cm,height=9cm}
\caption{Temperature dependence of the two length scales $\tilde{z}$ and $\xi_{\rm
max}$. The inset shows the ratio between these two quantities.}
\label{fig5}
\end{figure}

Finally we discuss the temperature dependence of the length scales
$\tilde{z}$ and $\xi_{\rm max}$. These quantities are shown in
figure~\ref{fig5} as a function of temperature. We see that both length
scales increase with decreasing temperature but that this increase is
relatively modest.  In the temperature interval $0.5 \leq T \leq 1.0$,
where the relaxation times of the system in the bulk increases by about a
factor of 500, this increase is only around 2.5 for the case of $\xi_{\rm
max}$ and a factor of 4 for $\tilde{z}$. Due to the slow relaxation of
the particles close to the wall it is presently not possible to study the
relaxation dynamics of these particles at significantly lower temperatures
and hence it is also not possible to determine neither $\tilde{z}$
nor $\xi_{\rm max}$ at low temperatures. Due to this relatively weak
increase of the length scales it is difficult to make precise statements
on whether or not these length scales diverge or not and if yes,
at which temperature this would be expected to happen. Furthermore it
is presently not even clear whether such a divergence occurs at the
same temperature, since the ratio $\tilde{z}/\xi_{\rm max}$ depends on
temperature (see inset of figure~\ref{fig5}). Since previous simulations
have identified two relevant temperatures, the critical temperature $T_c$
of mode-coupling theory at $T_c=0.435$ and the Kauzmann temperature
$T_K$ at around $T_K=0.29$ (Coluzzi {\it et al.} 2000a/b, Sciortino {\it et al.}
1999) we have tried to see whether the increasing
length scale is compatible with a power-law divergence at one of these
two temperatures. Whereas it seems that no such divergence occurs at
$T_c$, the data is indeed compatible with a critical behavior at $T_K$
with an exponent of the power-law around 1, both for $\tilde{z}$ and
$\xi_{\rm max}$. This result is in agreement with 
a theoretical approach to describe the temperature dependence of the size
of cooperatively rearranging region in the concept of configurational
entropy (Huth {\it et al.} 2000) . However, it is also possible
that the divergence occurs only at $T=0$ since also this temperature is
compatible with our data. Hence we conclude that although we are able
to identify a growing length scale it is presently not yet possible to
give a definite answer regarding the precise temperature dependence of
the scale. Therefore simulations at lower temperatures as well as some
theoretical guidance on this matter would be highly desirable.

{\bf Acknowledgements:} This work was supported by SFB 262/D1 of the
Deutsche Forschungsgemeinschaft.  We also thank the HLRZ J\"ulich for
a generous grant of computer time on the T3E.

\section*{References}
\begin{trivlist}
\item[]
Adams G., and Gibbs J.H., 1958,
{\it J. Chem. Phys.} {\bf 43}, 139.

\item[]
Cicerone M.T., Blackburn F.R., and Ediger M.D., 1995,
{\it Macromol.} {\bf 28}, 8224.

\item[]
Coluzzi B., Parisi G., and Verrocchio P., 2000a,
{\it J. Chem. Phys.} {\bf 112}, 2933.

\item[]
Coluzzi B., Parisi G., and Verrocchio P., 2000b,
{\it Phys. Rev. Lett.} {\bf 84}, 306.

\item[]
Doliwa B., and Heuer A., 2000,
{\it Phys. Rev. E}, {\bf 61}, 6898.

\item[]
Donati C., Glotzer S.C., Poole P.H., Kob W., and Plimpton S.J., 1999a,
{\it Phys. Rev. E}, {\bf 60}, 3107.

\item[]
Donati C., Glotzer S.C., and Poole P.H., 1999b,
{\it Phys. Rev. Lett.} {\bf 82}, 5064.

\item[]
Ediger M.D., 2000,
{\it Annual Rev. of Phys. Chem.} {\bf 51}, 99. 

\item[]
Franz S., and Parisi G., 1998,
cond-mat/9804084.

\item[]
Franz S., Donati C., Parisi G., and Glotzer S.C., 1999
{\it Phil. Mag. B}, {\bf 79}, 11.

\item[]
Franz S., and Parisi G., 2000,
{\it J. Phys.: Condens. Matter} {\bf 12}, 6335.

\item[]
Gleim T., Kob W., and Binder K., 1998,
{\it Phys. Rev. Lett.} {\bf 81}, 4404.

\item[]
G\"otze W., 1999,
J. Phys.: Condens. Matter {\bf 10}, A1.

\item[]
Hansen J.-P., and McDonald I.R., 1986,
{\it Theory of Simple Liquids}, Academic Press, London.
 
\item[]
Hempel E., Hempel G., Hensel A., Schick C., and Donth E., 2000
{\it J. Phys. Chem. B} {\bf 104}, 2460.

\item[]
Huth H., Beiner M., and Donth E., 2000
{\it Phys. Rev. B} {\bf 61}, 15092.

\item[]
Kob W., and Andersen H.C., 1995a,
{\it Phys. Rev. E}, {\bf 52}, 4134.

\item[]
Kob W., and Andersen H.C., 1995b,
{\it Phys. Rev. E}, {\bf 52}, 4626.

\item[]
Leheny R.L., Menon N., Nagel S.R., Price D.L., Suzuya K., and Thiyagarajan P., 1996,
{\it J. Chem. Phys.} {\bf 105}, 7783.

\item[]
Perera D.N., and Harrowell P., 1999,
{\it J. Chem. Phys.} {\bf 111}, 5441.

\item[]
Richert R., 1996,
{\it Phys. Rev. B}, {\bf 54}, 15762.

\item[]
Scheidler P., Kob W., and Binder K., 2000,
{\it Europhys. Lett.} {\bf 52}, 277.

\item[]
Scheidler P., 2001,
{\it PhD thesis}, University of Mainz.

\item[]
Sciortino F., Kob W., and Tartaglia P., 1999,
{\it Phys. Rev. Lett.} {\bf 83}, 3214.

\item[]
Sillescu H., 1999,
{\it J. Non-Cryst. Solids}, {\bf 243}, 81.

\item[]
Tracht U., Wilhelm M., Heuer A., Feng H., Schmidt-Rohr K., and Spiess H.W. 1998,
{\it Phys. Rev. Lett.} {\bf 81}, 2727.

\item[]
Yamamoto R., and Onuki A., 1998,
{\it Phys. Rev. E}, {\bf 58}, 3515.

\item[]
Yamamuro O., Tsukushi I., Lindquist A., Takahara S., Ishikawa M., and Matsuo T.,
1998,
{\it J. Phys. Chem. B} {\bf 102}, 1605.

\end{trivlist}

\newpage

\vfill

\end{document}